\pdfoutput=1

\documentclass[]{emulateapj}

\usepackage{amsmath,amssymb}

\DeclareMathOperator\erf{erf}

\newcommand{\intd}{ \, \mathrm{d} }
\newcommand{\derd}{ \mathrm{d} }

\newcommand{\water}{ \mathrm{H_{2}O} }
\newcommand{\co}{ \mathrm{CO} }
\newcommand{\cor}{ \mathrm{C/O} }

\newcommand{\uv}{ \mathrm{UV} }

\newcommand{\hy}{ \mathrm{H_{2}} }
\newcommand{\ice}{ \mathrm{ice}}

\begin{document}

\title{TRACING WATER VAPOR AND ICE DURING DUST GROWTH} 

\author{Sebastiaan Krijt\altaffilmark{1}, Fred J. Ciesla\altaffilmark{1}, and Edwin A. Bergin\altaffilmark{2}}
\email{skrijt@uchicago.edu}

\altaffiltext{1}{Department of the Geophysical Sciences, The University of Chicago, 5734 South Ellis Avenue, Chicago, IL 60637, USA}
\altaffiltext{2}{Department of Astronomy, University of Michigan, 500 Church Street, Ann Arbor, Michigan 48109, USA}

\begin{abstract} %max 250 words
The processes that govern the evolution of dust and water (in the form of vapor or ice) in protoplanetary disks are intimately connected. We have developed a model that simulates dust coagulation, dust dynamics (settling, turbulent mixing), vapor diffusion, and condensation/sublimation of volatiles onto grains in a vertical column of a protoplanetary disk. We employ the model to study how dust growth and dynamics influence the vertical distribution of water vapor and water ice in the region just outside the radial snowline. Our main finding is that coagulation (boosted by the enhanced stickiness of icy grains) and the ensuing vertical settling of solids results in water vapor being depleted, but not totally removed, from the region above the snowline on a timescale commensurate with the vertical turbulent mixing timescale. Depending on the strength of the turbulence and the temperature, the depletion can reach factors of up to ${\sim}50$ in the disk atmosphere. In our isothermal column, this vapor depletion results in the vertical snowline moving closer to the midplane (by up to 2 gas scale heights) and the gas-phase $\cor$ ratio above the vertical snowline increasing. Our findings illustrate the importance of dynamical effects and the need for understanding coevolutionary dynamics of gas and solids in planet-forming environments.

\end{abstract}


\section{Introduction}
Water plays a central and versatile role during planet formation. When present as vapor, it acts as a coolant for the gas and controls the chemistry of primitive materials by setting the oxidation state. In the form of ice, it represents a considerable fraction of the solid mass budget and has the potential to trap other, more volatile molecules in the form of amorphous ice. The presence of water on planets influences geological processes and is essential for creating habitable conditions. As such, understanding water's journey from the molecular cloud to the protoplanetary disk and its delivery onto planetary embryos are active areas of research \citep[e.g.,][]{ciesla2006,mandell2007,visser2011b,vandishoeck2014}.

During the protoplanetary disk phase, the outcome of dust coagulation (the first stage of planet formation) is influenced by how much water is present, mainly because frozen-out ice on grain surfaces is believed to enhance their stickiness and thus promote particle growth \citep[e.g.,][]{supulver1997,gundlach2014}. Models of planetesimal formation for many years have been finding that growth beyond a meter in size is hard to achieve \citep[see][for a recent review]{johansen2014}. Water ice plays an important role in three recently proposed solutions for crossing the meter-size barrier. For example, highly-porous grains can potentially grow rapidly enough to overcome radial drift \citep{okuzumi2012,kataoka2013c,krijt2015}, but this is only possible when the porous grains can survive high-velocity collisions, something only icy grains are potentially capable of accomplishing \citep{dominiktielens1997,wada2013,gundlach2014}. Close to the radial snowline, condensation itself can increase the sizes of individual dust grains significantly \citep{ros2013}. Finally, by freezing out onto grains water can also increase the local solid abundance by a factor of ${\sim}2$ \citep{lodders2003}, which helps to create conditions for triggering the streaming instability (SI), allowing the formation of clumps that can gravitationally collapse and form planetesimals almost instantaneously \citep{youdin2005,bai2010a,bai2010b,drazkowska2014b,carrera2015}. 

At the same time, the evolution of the solids influence water vapor in a number of ways. First, by dominating the disk's opacity at UV wavelengths, small dust grains play an important role in setting the temperature structure in the disk, influencing where in the disk water is present as ice or vapor \citep[e.g.,][]{min2011}. As such, assumptions about the dust population (its total mass, size distribution, and vertical and radial distributions) are an important part of interpreting observations of water lines and in connecting observed intensities to physical abundances \citep[e.g.,][]{hogerheijde2011,antonellini2015,antonellini2016,blevins2016}. Moreover, dust grain dynamics (e.g., radial drift, gravitational settling) are expected to alter the distribution of water in the radial \citep{cuzzi2004,ciesla2006,salyk2008} and vertical \citep{meijerink2009} direction. 

Starting with \citet{bergin2010, hogerheijde2011, du2015}, and now \citet{du2016}, we have increasingly clear evidence that we have a water problem: models predict too much water in the disk surface layers. In the work of \citet{du2015}, this was particularly acute in TW Hya between $5{-}10\mathrm{~AU}$, where the 179 micron water line was expected to shine out -- but was not detected. This observed under-abundance of water vapor can potentially be understood in terms of the so-called `cold finger' effect of \citet{stevenson1988}, but working in the vertical direction: water vapor is mixed from the vapor-rich atmosphere down to the disk midplane and freezes out onto large, settled grains that act as a sink because they cannot be lofted back up. As a result, vapor is removed from the disk's surface layers \citep{meijerink2009}. It is unclear however, how far this depletion can continue. For example, the efficiency of the vapor diffusion toward the midplane depends on the details of the local dust population \citep{monga2015}. Moreover, the removal of volatiles from the disk atmosphere is not necessarily a one-way process, as small grains capable of transporting ice back to the disk surface are continuously being produced in destructive collisions of larger bodies \citep{dullemonddominik2005}.

With the evolution of water and dust so intimately connected, we build on the work of \citet{krijtciesla2016} and set out to model the temporal evolution of the vertical distributions of solids, water vapor, and water ice in the region outside the radial snowline. The goal is to develop a framework in which the relevant processes can be simulated simultaneously and self-consistently, connecting the processes happening in the midplane (coagulation, fragmentation, freeze-out of volatiles) to those taking place in the disk atmosphere (sublimation and volatile release), allowing us to study the interplay between the dust and water evolution.

\section{Model}\label{sec:model}
For illustrative purposes, we will focus on an isothermal column in a typical protoplanetary nebula, while assuming a constant, vertically-integrated abundance of gas, water and refractory dust. Here, we describe our nebula model and summarize the physics behind the vertical motions of vapor and solids, dust coagulation, and water-dust interaction. In Sect. \ref{sec:methods}, we develop a numerical model for simulating these processes over relevant timescales.

\subsection{Disk structure}
Our nebula model is based on the Minimum Mass Solar Nebula \citep{weidenschilling1977,hayashi1981}, with a radial gas surface density profile that can be described as
\begin{equation}\label{eq:Sigma_g}
\Sigma_\mathrm{g}(r) = 2000\mathrm{~g~cm^{-2}} \left( \frac{r}{1\mathrm{~AU}} \right)^{-3/2}.
\end{equation}
The temperature is assumed to be constant in the vertical direction, and varies with radius as
\begin{equation}\label{eq:T}
T(r) = 280\mathrm{~K} \left( \frac{r}{1\mathrm{~AU}} \right)^{-1/2}.
\end{equation}
For these assumptions, the vertical profile of the gas density equals
\begin{equation}\label{eq:rho_g}
\rho_\mathrm{g}(z) = \frac{\Sigma_\mathrm{g}}{\sqrt{2\pi} h_\mathrm{g}} \exp\left\{ -  \frac{z^2}{2 h_\mathrm{g}^2} \right\},
\end{equation}
with $h_\mathrm{g} = c_s / \Omega$ the gas pressure scale-height,  $c_s = \sqrt{ k_\mathrm{B} T / m_\mathrm{g} }$ the sound-speed, $\Omega$ the Keplerian frequency, and $k_\mathrm{B}$ the Boltzmann constant. We set $m_\mathrm{g}=2.3\mathrm{~amu}$. A small fraction of the gas is water vapor, whose local mass density we denote as $\rho_\water$. While $\rho_\water \ll \rho_\mathrm{g}$, and neglecting advection, the diffusion equation for the water vapor is given as \citep{dubrulle1995}
\begin{equation}\label{eq:diffusion_eq}
\frac{\partial \rho_\water}{\partial t} = D_\mathrm{g}  \frac{\partial}{\partial z} \left( \rho_\mathrm{g}  \frac{\partial}{\partial z}  \frac{\rho_\water}{ \rho_\mathrm{g}} \right),
\end{equation}
where the turbulent gas diffusivity $D_\mathrm{g}$ is parametrized as \citep{shakura1973}
\begin{equation}\label{eq:D_g}
D_\mathrm{g} = \alpha c_s h_\mathrm{g},
\end{equation}
and does not depend on $z$ in the isothermal approximation. The vertical diffusivity implies a vertical mixing timescale of $t_D \sim h_\mathrm{g}^2 / D_\mathrm{g} \sim 1/\alpha \Omega$.

\subsection{Dust grain model}\label{sec:grainmodel}
\begin{figure*}[t]
\centering
\includegraphics[clip=,width=.80\linewidth]{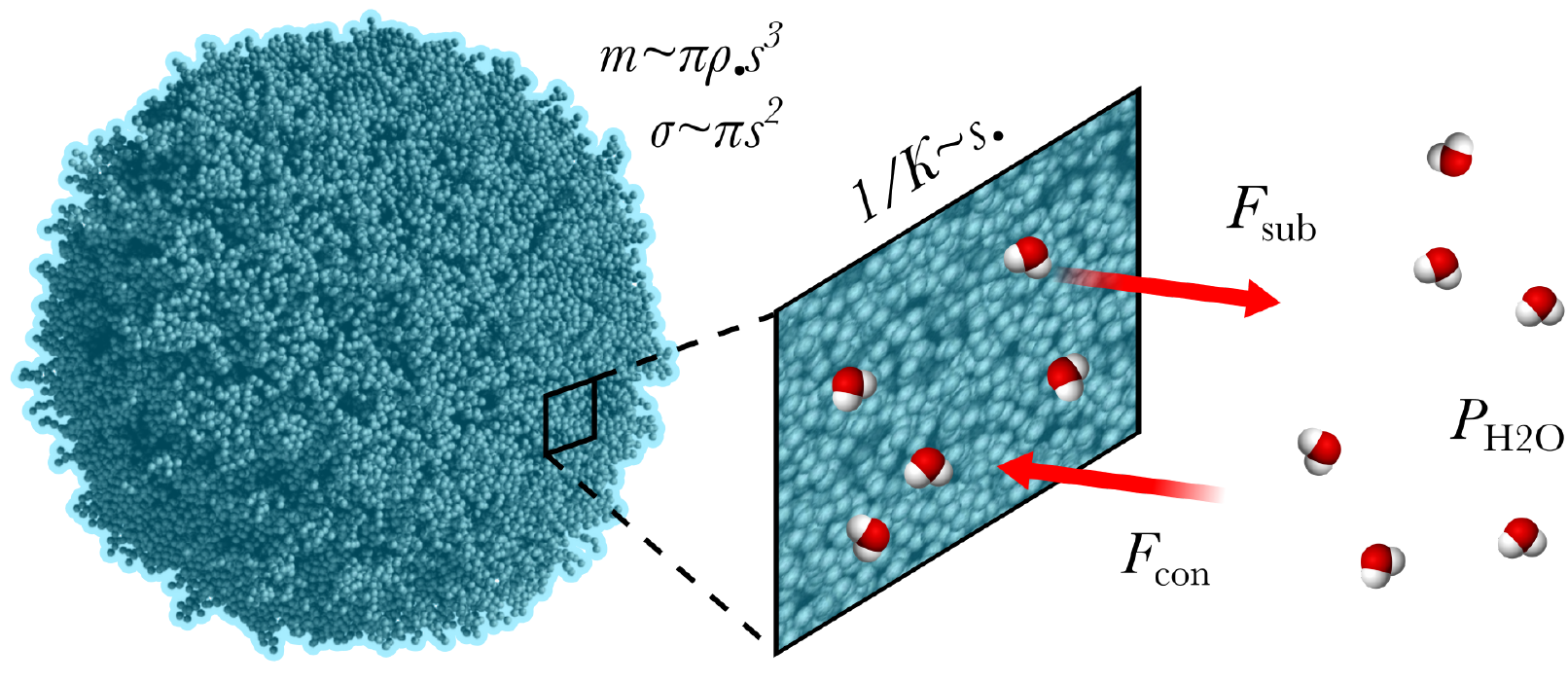}
\caption{Schematic of the dust particle model (not to scale). Aggregates, conglomerates of $\mathrm{\mu m}$-size monomers, are compact and have a fractal dimension close to 3, i.e., $m \sim \pi \rho_\bullet s^3$ (Sect. \ref{sec:grainmodel}). Their surface has a small-scale roughness with an average curvature $K\sim 1/s_\bullet$ reflecting the monomer size and exchanges water molecules with the gas via condensation and sublimation (Sect. \ref{sec:consub}). The fragmentation velocity of the aggregate is a function of its ice/rock ratio (Eqs. \ref{eq:f_ice} and \ref{eq:v_frag}). Aggregate image: \citet{seizinger2013c}.}
\label{fig:schematic_1}
\end{figure*}

Our initial dust population is assumed to be a mono-disperse distribution of refractory `monomers' with grain size $s_\bullet=1\mathrm{~ \mu m}$ and internal density $\rho_\bullet = 2.6 \mathrm{~g~cm^{-3}}$. The monomers can grow through collisions to form larger aggregates. While the (average) internal density of individual aggregates can in principle become much lower than that of the monomers themselves -- in particular when low-velocity collisions result in hit-and-stick growth \citep[e.g.,][]{kempf1999,ormel2007,okuzumi2012} -- we will assume in this work that growth leads to compact aggregates, with a fractal dimension close to 3, as depicted in Fig. \ref{fig:schematic_1}. In that case, an aggregate's mass $m$ and size $s$ are related through $m=(4/3) \pi s^3 \rho_\mathrm{int}$, where we will use $\rho_\mathrm{int} \approx \rho_\bullet$. The geometrical cross section of an aggregate is taken to equal $\sigma_\mathrm{geo} = \pi s^2$, and the surface area that is available for surface chemistry is written as $\sigma_\mathrm{chem} = 4 \pi s^2$.

Apart from its size, we will follow the amount of water ice on the aggregate as it moves through the disk. With aggregate-aggregate collisions, sublimation, and condensation all influencing the amount of ice, following how the ice is distributed on the aggregate is highly complex. Here, we only solve for the total number of water molecules on a grain, $n_\ice$, and assume the majority of the ice is concentrated close to the aggregate's surface. Furthermore, we assume the ice mantle does not significantly influence the grain's size, mass, or average density. The number of water molecules can be converted into a ice/rock mass ratio\footnote{We try to be consistent in using the subscript `H$_2$O' for water molecules in the gas phase, and `ice' for water molecules present on grains.}
\begin{equation}\label{eq:f_ice}
f_\ice \equiv \frac{m_\ice}{m_\mathrm{rock}} = \frac{ n_\ice m_\water}{(4/3) \pi s^3 \rho_\bullet},
\end{equation}
with $m_\water=18\mathrm{~amu}$ the mass of a single water molecule. The physical structure of a single dust aggregate is then fully characterized by its size (or mass) and ice fraction $f_\ice$ (or $n_\ice$). We return to the assumptions made here and their impact on the simulations in the remainder of this work in Sect. \ref{sec:disc}.

\subsubsection{Dynamics}
Including dust grain dynamics is vital for this study because ice-coated grains can transport water molecules between different regions of the disk. An important parameter for dynamics (and coagulation) is the particle's stopping time
\begin{equation}\label{eq:t_s}
t_s = \sqrt{\dfrac{\pi}{8}} \dfrac{\rho_\bullet s}{\rho_\mathrm{g} c_s} \left(1+\frac{4s}{9 \lambda_\mathrm{mfp}} \right),
\end{equation}
with $\lambda_\mathrm{mfp}= m_\mathrm{g} / (\sigma_\mathrm{mol} \rho_\mathrm{g})$ the molecular mean free path. It is common to define the dimensionless Stokes number $\mathrm{St}\equiv \Omega t_s$. We note that displacing a dust grain vertically changes its Stokes number and potentially even the drag regime the particle is in (Epstein or Stokes\footnote{The Epstein drag regime refers to the limit where $1 \gg 4s/9 \lambda_\mathrm{mfp}$ and the Stokes drag regime refers to $1 \ll 4s/9 \lambda_\mathrm{mfp}$.}) as both $\rho_\mathrm{g}$ and $\lambda_\mathrm{mfp}$ vary with height in the disk.

The vertical mixing of dust grains is then controlled by the dust particle diffusivity, related to $D_\mathrm{g}$ through the Schmidt number \citep{youdin2007}
\begin{equation}
\mathrm{Sc} \equiv \frac{D_\mathrm{g}}{D_\mathrm{d}} = 1 + (\Omega t_s)^2.
\end{equation}
For a given stopping time, an effective vertical velocity can be defined as
\begin{equation}\label{eq:v_eff}
v_\mathrm{eff} =  - D_\mathrm{d} \frac{z}{h_\mathrm{g}^2}  - \frac{t_s \Omega^2 z}{1+\Omega t_s}.
\end{equation}
which can be used to calculate the trajectory of an individual grain \citep[e.g.,][]{ciesla2010}. The factor $(1+\Omega t_s)^{-1}$ in the second term on the RHS of Eq. \ref{eq:v_eff} is there to ensure that the vertical velocity does not exceed the vertical component of the Kepler velocity for large grains with $\Omega t_s \gg 1$ \citep{brauer2007}. In the absence of collisions, the dependence of $v_\mathrm{eff}$ on particle size (through the stopping time) will result in large grains settling to the midplane while small grains (those with $\Omega t_s \ll \alpha$) keep a vertical distribution very similar to that of the gas \citep{dullemond2004,ciesla2010}. This picture can change when collisions are frequent, in which case small grains, after being released in collisions of larger bodies, can become trapped in the disk midplane \citep{krijtciesla2016}.

\subsubsection{Collisions}
Depending on the velocity at which a grain-grain collision occurs, grains can stick together or be (partially) destroyed \citep[e.g.][]{blumwurm2008,guttler2010}. When calculating relative collision velocities, we take into account Brownian motion, turbulence \citep{ormel2007b}, and differential settling, and include the dependence of the various velocity sources on the location, $z$, at which the collision takes place.

The maximum velocity at which collisions can result in growth is denoted by the fragmentation velocity $v_\mathrm{frag}$. We use a variable fragmentation threshold to capture the effect frozen-out water ice has on the stickiness of the grains, writing
\begin{equation}\label{eq:v_frag}
v_\mathrm{frag}=
\begin{cases}
~ 1\mathrm{~m/s} & \textrm{for~~}  f_\ice < f^*, \vspace{3mm}\\
~ 10\mathrm{~m/s} & \textrm{for~~}  f_\ice \geq f^*.
\end{cases}
\end{equation}
The factor of 10 difference between refractory and icy grains is supported by both numerical simulations \citep{dominiktielens1997,wada2013} and experimental results \citep{gundlach2014}. The threshold value for $f^*$ is less well constrained by theory or experiments. The sticking of microscopic spheres involves only a surface layer with thickness $\delta / s \sim 10^{-2}$ \citep[e.g.,][]{chokshi1993,krijt2013}, indicating a relatively small amount of water ice can be sufficient to alter the sticking properties (i.e., $f^* \ll 1$). For a larger aggregate however, the details will depend on how exactly the water ice is distributed. Here, we will use $f^*=0.1$, and note that the main results are not very sensitive to this choice as long as $f^* \lesssim 1$.

\subsection{Condensation/sublimation}\label{sec:consub}
The saturated vapor pressure for water on a flat surface equals \citep{supulver2000}
\begin{align}\label{eq:P_sat}
P_\mathrm{sat} &=  \dfrac{k_\mathrm{B} T}{m_\water}  \rho_\mathrm{sat}, \\ 
& = 1.013\times10^6  \exp \left\{ 15.6 - \dfrac{5940}{T} \right\}  \mathrm{~dyn/cm^2}.
\end{align}
On a non-flat surface however, this expression has to be modified to include the effect of surface curvature. Following \citet{sirono2011}, we write
\begin{equation}\label{eq:P_sat_K}
P^K_\mathrm{sat} = P_\mathrm{sat} (1+ K \gamma v / k_\mathrm{B} T),
\end{equation}
with $\gamma=69\mathrm{~erg/cm^2}$ the surface energy of water ice, $v=3.3\times10^{-23}\mathrm{~cm^3}$ the volume of a water molecule, and $K$ represents the local radius of curvature. With our aggregates being conglomerates of micron-size grains, we will assume that all aggregates, independent of their macroscopic size, have a characteristic (i.e., averaged over their entire surface) local curvature of $K = 1/ s_\bullet$ (see Fig. \ref{fig:schematic_1}). In addition, we assume that this $K$ is not influenced by the presence of water ice. For these parameters, typically $K \gamma v / k_\mathrm{B} T \sim 10^{-2}$ and $P^K_\mathrm{sat} \approx P_\mathrm{sat}$. The assumptions made here ignore some interesting effects, which we discuss in detail in Sect. \ref{sec:disc}.

For a given vapor pressure, the sublimation and condensation rates (in $\mathrm{g~cm^{-2}~s^{-1}}$) are given by \citep[e.g.,][]{supulver2000}
\begin{equation}\label{eq:F_sub}
F_\mathrm{sub} = - \sqrt{\frac{m_\water}{2\pi k_\mathrm{B} T }} P^K_\mathrm{sat},
\end{equation}
\begin{equation}\label{eq:F_con}
F_\mathrm{con} =  \sqrt{\frac{m_\water}{2\pi k_\mathrm{B} T }} P_\water,
\end{equation}
which we can combine to find the rate of change of the total number of water molecules on a given grain as
\begin{equation}\label{eq:dndt}
\frac{\derd n_\ice}{\derd t} =  \frac{ \sigma_\mathrm{chem}}{m_\water}  \frac{v_\mathrm{th}}{4}  (\rho_\water - \rho^K_\mathrm{sat}),
\end{equation}
where we have written $v_\mathrm{th}=(8 k_\mathrm{B} T / \pi m_\water)^{1/2}$ as the thermal velocity of water molecules. The RHS of Eq. \ref{eq:dndt} is positive when $P_\water > P^K_\mathrm{sat}$ (condensation dominates) and negative when $P_\water < P^K_\mathrm{sat}$ (sublimation dominates).

\section{Numerical methodology}\label{sec:methods}
In this Section we develop a framework that allows us to simultaneously model the physical processes described in Sect. \ref{sec:model}. The model builds on the method of \citet{krijtciesla2016}, where coagulation and vertical mixing of dust were combined. Apart from the addition of the water vapor, an important difference with that work is that we now solve for the dust size distribution self-consistently, rather than assuming a coagulation/fragmentation steady state from the start.

\subsection{Set-up and initial conditions}\label{sec:IC}
The simulated region is a single vertical column at a radius $r$ from a sun-like star, and the vertically integrated refractory dust and water+ice contents of the column are taken to equal $\Sigma_\mathrm{d}/\Sigma_\mathrm{g}=5\times10^{-3}$ and $\Sigma_{\water+\ice}/\Sigma_\mathrm{g}=5\times10^{-3}$ \citep[e.g.,][]{lodders2003}. The column is described by a total of $N_\mathrm{b}$ stacked boxes placed between $z=0$ and $z=4h_\mathrm{g}$ with heights $\mathcal{L} = 4 h_\mathrm{g}/N_\mathrm{b}$ and volumes $\mathcal{V}=\mathcal{L}^3$. Inside a single box, the temperature, gas density, and vapor density are assumed to be uniform, and the dust particles inside the box are treated as being well-mixed spatially. The dust population in the column is represented by $N_\mathrm{p}$ super-particles. Every super-particle $i$ represents $\mathcal{N}_i$ physical particles that together have a mass $M_i$. We choose to keep $M_i$ fixed for all super-particles, so that $\mathcal{N}_i = M_i / m_i$, with $m_i$ the mass of an individual particle. The total mass of all physical particles must equal 50\% of the total dust mass in the column\footnote{The 50\% comes from the fact we are only simulating the upper half of the disk.}, i.e., $N_\mathrm{p} M_i = \Sigma_\mathrm{d}(r) \mathcal{L}^2 /2$. Gas-phase water molecules are not followed individually, but we solve for the water vapor density $\rho_\water(t)$ inside every box taking into account diffusion between neighboring boxes and losses/gains through condensation/sublimation, while assuming that the bulk gas density and temperature are not influenced by the water vapor and dust distributions.

At $t=0$, before we start the calculation, we distribute the refractory dust and water vapor such that it is well-mixed with the gas. The $N_\mathrm{p}$ refractory dust particles start out as monomers with radii $s_i = s_\bullet=1\mathrm{~\mu m}$ and initial $z_i \geq 0$ are drawn from a Gaussian with half-width $h_\mathrm{g}$, resulting in $(\rho_\mathrm{d} / \rho_\mathrm{g}) \approx \Sigma_\mathrm{d}/\Sigma_\mathrm{g} = 5\times10^{-3}$ inside every cell\footnote{We take care to ensure that there is at least 1 super-particle in every grid cel at $t=0$.}. The water vapor is first distributed according to $(\rho_\water / \rho_\mathrm{g}) = \Sigma_{\water}/\Sigma_\mathrm{g} = 5\times10^{-3}$, and then (without being allowed to diffuse) given the opportunity to equilibrate by freezing out onto the present dust grains. In the region just beyond the radial midplane snowline ($r\simeq 3.2\mathrm{~AU}$ in our disk model), this results in two distinct vertical regions in the isothermal approximation. Due to low gas densities ($\rho_\water < \rho^K_\mathrm{sat}$), the water stays in the vapor phase at high $z$, resulting in a region of constant water-vapor abundance and grains with $f_\ice=0$. Closer to the midplane, the original well-mixed water vapor abundance exceeds $\rho^K_\mathrm{sat}$ and the molecules condense onto the dust grains. This results in a region where $\rho_\water = \rho^K_\mathrm{sat}$, leading to a decreasing $n_\water/n_\hy$ with decreasing $z$ (since $\rho^K_\mathrm{sat}$ is constant) and grains covered in ice. We will refer to the boundary between these regions as the \emph{vertical snowline}. The (initial) location of this line can be found by solving $\rho_\water = \rho^K_\mathrm{sat}$, resulting in
\begin{equation}\label{eq:ZSL_0}
z_\mathrm{SL}^{0} = h_\mathrm{g} \left[ -2 \ln \left( \frac{\sqrt{2 \pi} m_\water h_\mathrm{g} P^K_\mathrm{sat} }{ \Sigma_\water k_\mathrm{B} T} \right)    \right]^{1/2},
\end{equation}
where $P^K_\mathrm{sat}$ is a function of temperature (Eq. \ref{eq:P_sat_K}). In the model outlined here (e.g., vertically isothermal), the vertical snowline will remain at $z_\mathrm{SL}^{0}$ when vapor diffusion and vertical motions of dust grains are ignored (even if growth is taking place). As such, we can use this `static' limit as a reference point for comparing simulation results.

The following Sections detail how the dust and vapor are evolved in time. Specifically, we discuss how coagulation (Sect. \ref{sec:dustgrowth}) and gas-grain chemistry (Sect. \ref{sec:chemistry}) are calculated per box, and how dust dynamics (Sect. \ref{sec:dynamics}) and vapor diffusion (\ref{sec:diffusion}) allow dust and vapor to move from one box to another.

\subsection{Dust dynamics}\label{sec:dynamics}
Grain dynamics (gravitational settling and turbulent vertical diffusion) are calculated according to \citet{ciesla2010}, in which a particle's vertical location after some time $
\Delta t$ is obtained as
\begin{equation}\label{eq:z_new}
z_i(t+\Delta t) = z_i(t) + v_\mathrm{eff} \Delta t + \mathcal{R}_1 \left[ \frac{2}{\zeta} D_\mathrm{d} \Delta t \right]^{1/2},
\end{equation}
with $\zeta=1/3$ and $\mathcal{R}_1$ is a random number between $[-1,1]$ and the effective velocity given by Eq. \ref{eq:v_eff} depends on the location and size of the grain. Based on the dynamics, the global time-step $\Delta t_\mathrm{global}$ is chosen such that for all super-particles
\begin{equation}\label{eq:dt_global}
| z_i(t+\Delta t_\mathrm{global}) - z_i(t) | < \mathcal{L},
\end{equation}
to ensure that grains do not traverse multiple box boundaries in a single time-step. It is this global timestep that we base the rest of our calculations on.

\subsection{Dust growth and fragmentation}\label{sec:dustgrowth}
During a global timestep $\Delta t_\mathrm{global}$, coagulation and fragmentation are solved per box, using the representative particle approach of \citet{zsom2008} and \citet{zsom2011}. In short, the method comes down to calculating all collision rates between pairs of particles present inside that particular box, and using random numbers to determine which particles collide and when. In this way, we calculate forward in time from collision to collision, until a period $\Delta t_\mathrm{global}$ has passed and we allow grains to move between adjacent boxes again.

The collision rate of super-particle $i$ with a particle represented by super-particle $j$ is
\begin{equation}
C_{ij} = 
\begin{cases}
~n_j  v_\mathrm{rel}  \sigma_{ij} & \textrm{if $i$ and $j$ in same box,}\\
~0 & \textrm{if $i$ and $j$ in different boxes,}
\end{cases}
\end{equation}
with $n_j = \mathcal{N}_j / \mathcal{V}$ the number density of $j$ particles, $v_\mathrm{rel}$ the relative velocity between those particles, and $\sigma_{ij} = \pi( s_i + s_j)^2$ the geometric collisional cross section. For numerical reasons, we modify the collision rate when $m_j < f_\epsilon m_i$, in which case we group $j$-particles together and collisions occur between a single $i$-particle and a total of $f_\epsilon m_i / m_j)$ identical $j$-particles \citep[see][]{krijtciesla2016}. The modified collision rates read
\begin{equation}\label{eq:C_tilde}
\widetilde{C}_{ij} = \begin{cases}
~C_{ij} &\textrm{~~if~~} (s_j/s_i)^3 > f_\epsilon,  \vspace{3mm}\\
~f_\epsilon^{-1} (s_j/s_i)^3 \cdot C_{ij} &\textrm{~~if~~} (s_j/s_i)^3 \leq f_\epsilon,
\end{cases}
\end{equation}
and we use $f_\epsilon=0.1$.

To solve coagulation in a single box, we first compute the collision rates for all super-particles inside that box, and sum over the individual contributions to find 
\begin{equation}\label{eq:C_i}
\widetilde{C}_i = \sum_j \widetilde{C}_{ij},
\end{equation}
and
\begin{equation}\label{eq:C_tot}
\widetilde{C}_\mathrm{tot} = \sum_i \widetilde{C}_{i}.
\end{equation}
A random number $\mathcal{R}$ is drawn from a uniform distribution between 0 and 1, and the time until the next collision event is computed as
\begin{equation}
\Delta t_\mathrm{col} = - \frac{\ln \mathcal{R}}{\widetilde{C}_\mathrm{tot}}.
\end{equation}
If $\Delta t_\mathrm{col} > \Delta t_\mathrm{global}$, no collision event takes place in this particular box during the global time-step, and we move on to the next box. Alternatively, when $\Delta t_\mathrm{col} \leq \Delta t_\mathrm{global}$, we use two more random numbers to identify the collision partners $i$ and $j$, and another random number to determine the collisional outcome for the corresponding fragmentation probability. In the case of fragmentation, we need one more random number to choose the resulting fragment size (see Eq. 13 of \citet{krijtciesla2016} for the fragment size distribution). This process is repeated until the sum of $\Delta t_\mathrm{col}$ exceeds $\Delta t_\mathrm{global}$, in other words, multiple collisions can occur within the same box in a given time-step \citep[see also][]{zsom2011}. We note that the individual collision rates (and therefore also the sums of Eqs. \ref{eq:C_i} and \ref{eq:C_tot}) have to be re-evaluated after every single collision event because the properties of super-particle $i$ have been altered.

We employ the collisional outcome model of \citet{birnstiel2011}, in which the fragmentation probability $P_\mathrm{frag}$ is 1 at collision velocities above $v_\mathrm{frag}$, 0 below $0.8 v_\mathrm{frag}$, and has a linear transition between these two regimes. In this collision model bouncing is neglected, so that the probability of sticking equals $P_\mathrm{stick} = 1- P_\mathrm{frag}$.

\subsection{Sublimation and condensation}\label{sec:chemistry}
The amount of water vapor inside a given box will change as the result of sublimation/condensation, as dust grains return/remove water molecules to/from the surrounding gas.

While $\rho_\water > \rho^K_\mathrm{sat}$ inside a box, the change in vapor density following from condensation can be written as
\begin{equation}
\frac{\derd \rho_\water}{\derd t} = - \frac{m_\water}{\mathcal{V}} \sum_{i} \frac{M_i}{m_i} \left(\frac{\derd n_\ice}{\derd t }\right)_i,
\end{equation}
where the sum is over all $i$ super-particles that are present inside that particular box. The fraction $M_i/m_i$ takes into account that super-particle $i$ represents $\mathcal{N}_i$ physical particles. Inserting Eq. \ref{eq:dndt} results in
\begin{equation}
\frac{\derd \rho_\water}{\derd t} = - (\rho_\water
 - \rho^K_\mathrm{sat}) \underbrace{ \frac{3 v_\mathrm{th} }{4 \mathcal{V} \rho_\bullet}  \sum_i M_i s_i^{-1}}_{ A },
\end{equation}
where $A$ does not depend on $\rho_\water$. Defining $\rho^* = (\rho_\water - \rho^K_\mathrm{sat})$, we are left with
\begin{equation}
\frac{\derd \rho^*}{\derd t} = \frac{\derd \rho_\water}{\derd t} = - A \rho^*, 
\end{equation}
leading to 
\begin{equation}\label{eq:condens}
\rho^*(t+\Delta t) =\rho^*(t)  \exp(-A \Delta t).
\end{equation}
Thus, to calculate the time-dependent condensation during a global timestep $\Delta t_\mathrm{global}$ we only need to evaluate $A$ (which requires summing over the super-particles inside box $b$) once, before making use of Eq. \ref{eq:condens}. After computing how much water vapor is actually lost from the gas during $\Delta t_\mathrm{global}$, we distribute these water molecules to the present super-particles in a cross-section-weighted way, i.e., proportional to $s^2$ (see Eq. \ref{eq:dndt}).

Sublimation is handled slightly differently. Because grains only have a finite number of molecules to give back to the gas, Eq. \ref{eq:dndt} is solved for each super-particle in the box independently, while the change in ice molecules is limited to ensure $n_\ice(t+\Delta t) \geq 0$. The new $\rho_\water(t+\Delta t)$ is computed at end of the timestep by adding the contributions of the relevant super particles.

\subsection{Water vapor diffusion}\label{sec:diffusion}
Lastly, we simulate vertical vapor diffusion during a time-step $\Delta t_\mathrm{global}$ by solving Eq. \ref{eq:diffusion_eq} using the method of finite differences. For the boundary conditions, we assume that there is no vapor flux through the top of the upper box and the bottom of the lower box (the disk midplane).

\subsection{Final recipe}
The recipe for evolving the column then becomes:
\begin{enumerate}
\item{Determine global time-step $\Delta t_\mathrm{global}$;}
\item{For every box, solve coagulation/fragmentation until $\Delta t_\mathrm{global}$ is reached;}
\item{Displace all dust particles vertically;}
\item{Calculate sublimation and condensation inside every box;}
\item{Allow water vapor to diffuse between adjacent boxes.}
\item{Repeat steps 1--5.}
\end{enumerate}
Only steps 2 (if a collision occurs) and 3 (if a grain moves from one box to another) require that the collision rates are updated. Because we assume that the water-ice fraction of the grains does not influence their size or Stokes number, and that the vapor distribution does not impact the bulk gas properties, the collision rates stay the same during all other computational steps.

\subsubsection{Optical depths}\label{sec:tau}
While not used directly in the simulations, an interesting output of the models is the vertical profile of the dust optical depth. In general, the opacity, $\kappa_\lambda$, of a dust grain depends on the size, shape, and composition of the grain, as well as the wavelength $\lambda$ \citep[e.g.,][]{kataoka2014,cuzzi2014}. For the particle sizes considered in this work, however, an adequate approximation at UV and visible wavelengths is that the optical cross section equals the geometrical cross section, resulting in $\kappa_\lambda \sim1/s_i \rho_\bullet$. The optical depth of a group of grains represented by a single representative particle can then be written as
\begin{equation}\label{eq:tau_i}
\tau_i = \frac{\pi s_i^2}{\mathcal{L}^2} \mathcal{N}_i = \frac{3}{8} \frac{\Sigma_\mathrm{d}}{s_i \rho_\bullet} N_\mathrm{p}^{-1},
\end{equation}
and the cumulative vertical UV optical depth at height $z$, measured from above (i.e., perpendicular to the midplane), becomes
\begin{equation}
\tau_\uv(z) =  \sum_{i,z_i > z} \tau_i.
\end{equation}
An important location is the height where $\tau_\uv(z)=\mu_0$ (with $\mu_0 \sim h_\mathrm{g}/r \sim 0.05$ the flaring angle of the disk), which indicates the height below which stellar photons cannot efficiently penetrate \citep{dalessio1998}. For example, dust grains in the region where $\tau_\uv(z) > \mu_0$ will receive considerably less of the incident stellar UV radiation, protecting them from photo-desorption (see Sect. \ref{sec:PD}). For the conditions at $t=0$ (all grains are $s_\bullet$ in size and the dust-to-gas ratio is constant vertically), we can obtain
\begin{equation}\label{eq:ZTAU_0}
z_{\tau_\uv=\mu_0}^0 = \sqrt{2} h_\mathrm{g} \erf^{-1} \left(1- \mu_0 \frac{8s_\bullet \rho_\bullet}{3 \Sigma_\mathrm{d}} \right),
\end{equation}
with $\erf^{-1}$ the inverse error function. As coagulation and settling of dust begin to take place, the location where $\tau_\uv(z)=\mu_0$ is expected to move closer to the midplane.

%At longer wavelengths, the $\tau_\lambda=1$ surface lies closer to the midplane, with the dust disk typically being optically thin at mm wavelengths \citep[e.g.,][]{testi2014}.}

\subsubsection{Choosing $N_\mathrm{p}$ and $N_\mathrm{b}$}\label{sec:resolution}
The computational cost involved with these calculations depends heavily on the choice of the number of super-particles $N_\mathrm{p}$ and the number $N_\mathrm{b}$ of grid cells. To accurately capture the vertical variation of the dust density after grain settling, the size of single cell should be smaller than the scale-height of the largest grains, which can become problematic if settling is very effective \citep[see Sect. 4 of][]{drazkowska2013}. In our case, fragmentation and turbulent mixing will limit settling, and it suffices to use 20 cells per gas scale-height (i.e., $\mathcal{L}=0.05h_\mathrm{g}$) for the $\alpha=10^{-3}$ case and twice as many for the $\alpha=5\times10^{-4}$ runs.

The number of super-particles also cannot be chosen to be too small \citep[see also][]{zsom2008}. In particular, it is important that we resolve the particle size distribution and the vertical profile of the dust. Moreover, to resolve $z_{\tau_\uv=\mu_0}$, we need to make sure a single super-particle is sufficiently optically thin, i.e., $\tau_i < \mu_0$. Making use of Eq. \ref{eq:tau_i}, we find $N_\mathrm{p} \gtrsim 10^4$ for $s_\bullet = 1 \mathrm{~\mu m}$ and $\Sigma_\mathrm{d}=1\mathrm{~g/cm^2}$. Based on these considerations, we will use a $10^4 < N_\mathrm{p} < 10^5$ in the remainder of this work.

%First, we want to resolve the vertically integrated size distribution

%For the simulations presented here, the typical abundance (in terms of fraction of the total dust mass) of small grains is of the order of $10^{-2}$. Having a couple of small grains is not enough though, since we want to accurately resolve their vertical distribution. Finally, 

%At $t=0$, on average 1 in every ${\sim}16,000$ monomers will be located above $4 h_\mathrm{g}$, so we need $N_\mathrm{p} \gtrsim 10^{4}$ to have at least some dust in the upper-most boxes.

\subsubsection{Conservation of total dust and water mass}
Since we are looking at a closed column of gas and dust, the total mass of refractory particles and water molecules (present as ice or vapor) should be conserved. In the method outlined in this section, this is true for the mass in refractory grains, but not necessarily for the water molecules. While there is no net flux of water vapor through the bottom and top boundaries of the column, the mass in water ice can vary because of how the representative particle approach works. Specifically, when a representative grain collides, only the properties of \emph{that} representative particle are changed, while the properties of the grain representing the particle it collides with are left unaltered \citep{zsom2008,zsom2011}. This causes fluctuations in the total water ice mass when collisions occur between grains of very dissimilar ice-to-rock ratios. For the simulations presented in this paper, the fractional variations, measured between start and end of the simulations, were typically ${\sim}10^{-3}$, and up to factor 10 larger for the simulations at $r=3.5\mathrm{~AU}$, where collisions between grains with very different $f_\ice$ are most common. Thus, small changes in the $\water$ abundance are seen, but they are small compared to the effects modeled here.

\begin{figure*}[t]
\centering
\includegraphics[clip=,width=.95\linewidth]{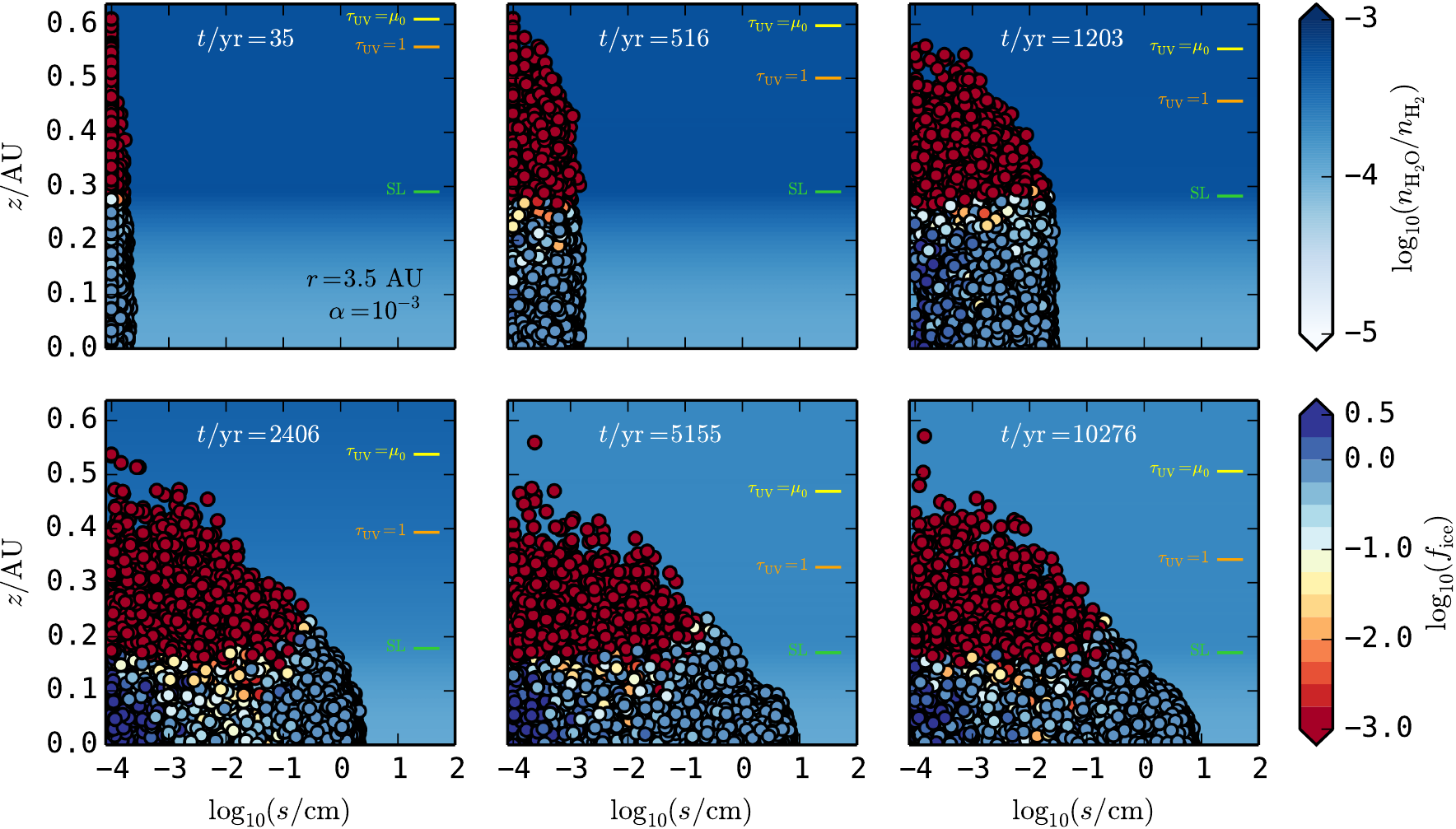}
\caption{Dust, ice, and water vapor evolution at $r=3.5\mathrm{~AU}$ for $\alpha=10^{-3}$ simulated over $10$ mixing times $t_D$. The circles indicate sizes and locations of representative particles, with the colors showing ice-to-rock mass ratio. The background color indicates the water vapor abundance relative to $\hy$. The horizontal markers indicate the location of the vertical snowline and the heights where the cumulative optical depth exceeds $\mu_0$ or $1$ (see text).}
\label{fig:35AU_NT}
\end{figure*}

\section{Results}\label{sec:results}
Here we use the methodology outlined in Sect. \ref{sec:methods} to model the evolution of dust, ice, and water vapor in isolated columns at radii between 3 and $4.5\mathrm{~AU}$. At every radial location, the gas surface density and (vertically isothermal) temperature are given by Eqs. \ref{eq:Sigma_g} and \ref{eq:T}. The simulations together cover both sides of the radial snowline, which is located at $r\simeq 3.2\mathrm{~AU}$ for our disk model. 

%We first discuss a couple of individual simulations in detail, and then study how the observed behavior changes as a function of disk radius and strength of the turbulence.

\begin{figure*}[t]
\centering
\includegraphics[clip=,width=.95\linewidth]{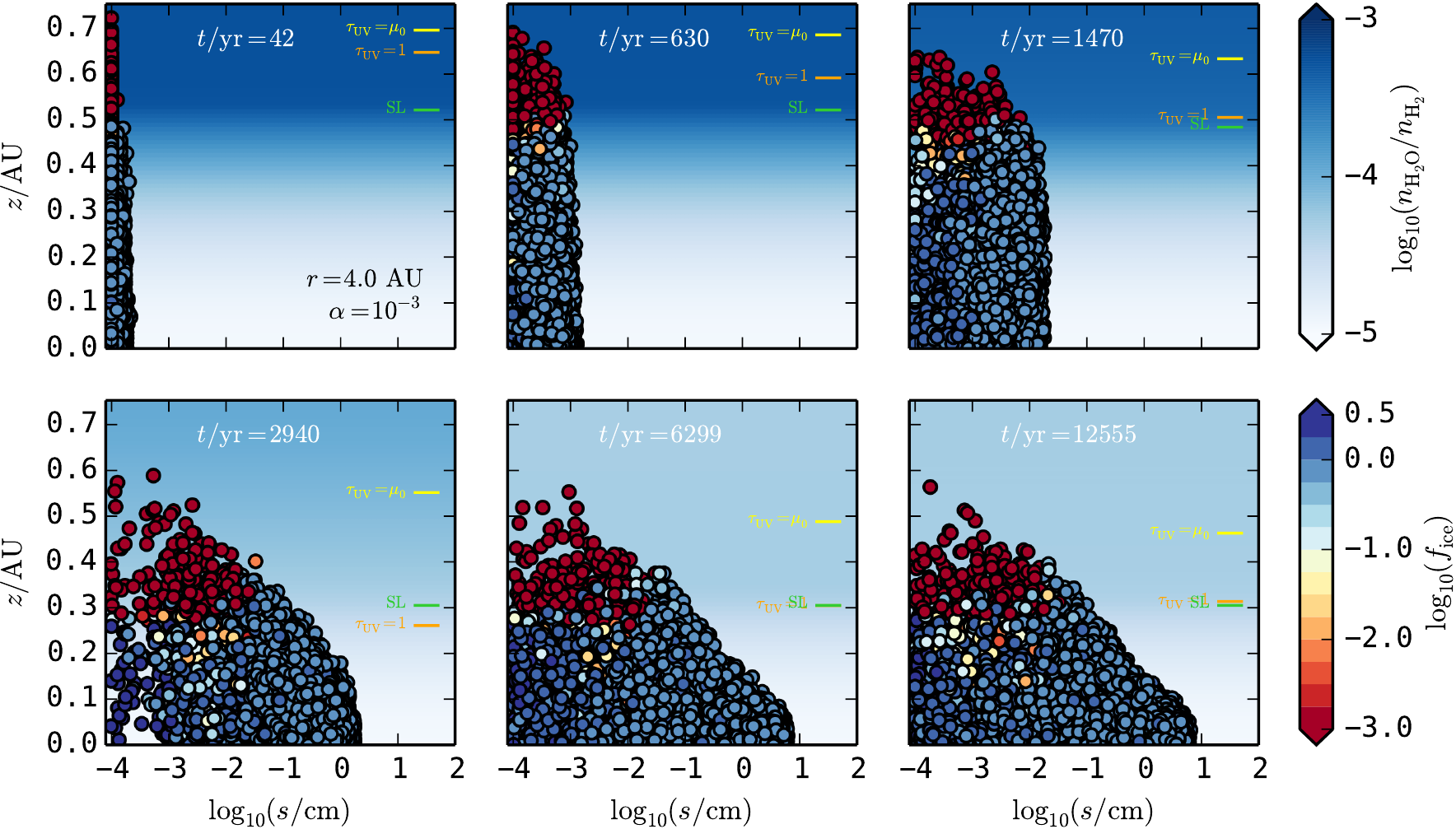}
\caption{Same as Fig. \ref{fig:35AU_NT}, but at $r=4\mathrm{~AU}$. The lower temperature (Eq. \ref{eq:T}) results in the vertical snowline being located further from the midplane.}
\label{fig:4AU_NT}
\end{figure*}

%The lower temperature and gas surface density result in the snowline being located higher up in the disk. The $\tau_\uv=1$ surface (purple marker) beings very high in the disk when all grains are small, reaches a minimum after ${\sim}3300\mathrm{~yr}$ when the size-distribution is fairly narrow, and comes back up to ${\sim}0.3\mathrm{~AU}$ after $6000\mathrm{~yr}$ when the population of small grains is replenished by fragmentation.

Figures \ref{fig:35AU_NT} and \ref{fig:4AU_NT} show the evolution of the dust and water (vapor and ice) over 10 mixing times ($t_D\sim10^3\mathrm{~yr}$ for this model, see Table \ref{tab:results}) in columns at $r=3.5\mathrm{~AU}$ and $4\mathrm{~AU}$ for a turbulence characterized by $\alpha=10^{-3}$. The spherical symbols show the locations, size, and ice/rock ratio of the dust super-particles (Eq. \ref{eq:f_ice}), and the background color shows the water vapor abundance relative to $\hy$, calculated as
\begin{equation}
\frac{n_\water}{n_\hy} = \frac{m_\hy}{m_\water} \frac{\rho_\water}{ \rho_\mathrm{g} } = \frac{1}{9} \frac{\rho_\water}{ \rho_\mathrm{g} }.
\end{equation}
The green marker shows the location of the vertical snowline, defined as the location where the water vapor density drops below $99\%$ of $\rho^K_\mathrm{sat}$, and the yellow and orange marker show where the cumulative optical depth, as calculated from above, reaches $\mu_0$ and $1$, respectively. The vertical extent of the plot corresponds to $4 h_\mathrm{g}$. The top-left panel of both figures closely resemble the initial conditions (see Sect. \ref{sec:IC}): we have well-mixed water vapor and ice-poor grains above the snowline, and ice-rich grains and a water abundance relative to $\hy$ that decreases with decreasing $z$ below the snowline. The initial locations of the snowlines in both figures are in good agreement with Eq. \ref{eq:ZSL_0} when we insert $\Sigma_\water/\Sigma_\mathrm{g}=5\times10^{-3}$ (see also Table \ref{tab:results}).

Focusing first on the time evolution of the solid component, we see that the dust grains grow to ${\sim}\mathrm{cm}$ sizes on timescales of several thousand years and in general, grains below the snowline are ice-rich while grains in the upper parts of the disk are ice-poor. Looking at the vertical locations of the various representative particles, it is clear that gravitational settling is an important effect for grains larger than a millimeter or so, which is approximately the size for which $\mathrm{St} \sim \alpha$ \citep[e.g.,][]{krijtciesla2016}. In the next Sections, we discuss in detail the resulting dust \& ice distributions (Sects. \ref{sec:nama} and \ref{sec:icerock}) and the effect grain growth and settling has on the vapor content of the disk atmosphere (Sect. \ref{sec:depletion}) and the location of the vertical snowline (Sect. \ref{sec:location}).

\begin{figure}[!t]
\centering
\includegraphics[clip=,width=.95\linewidth]{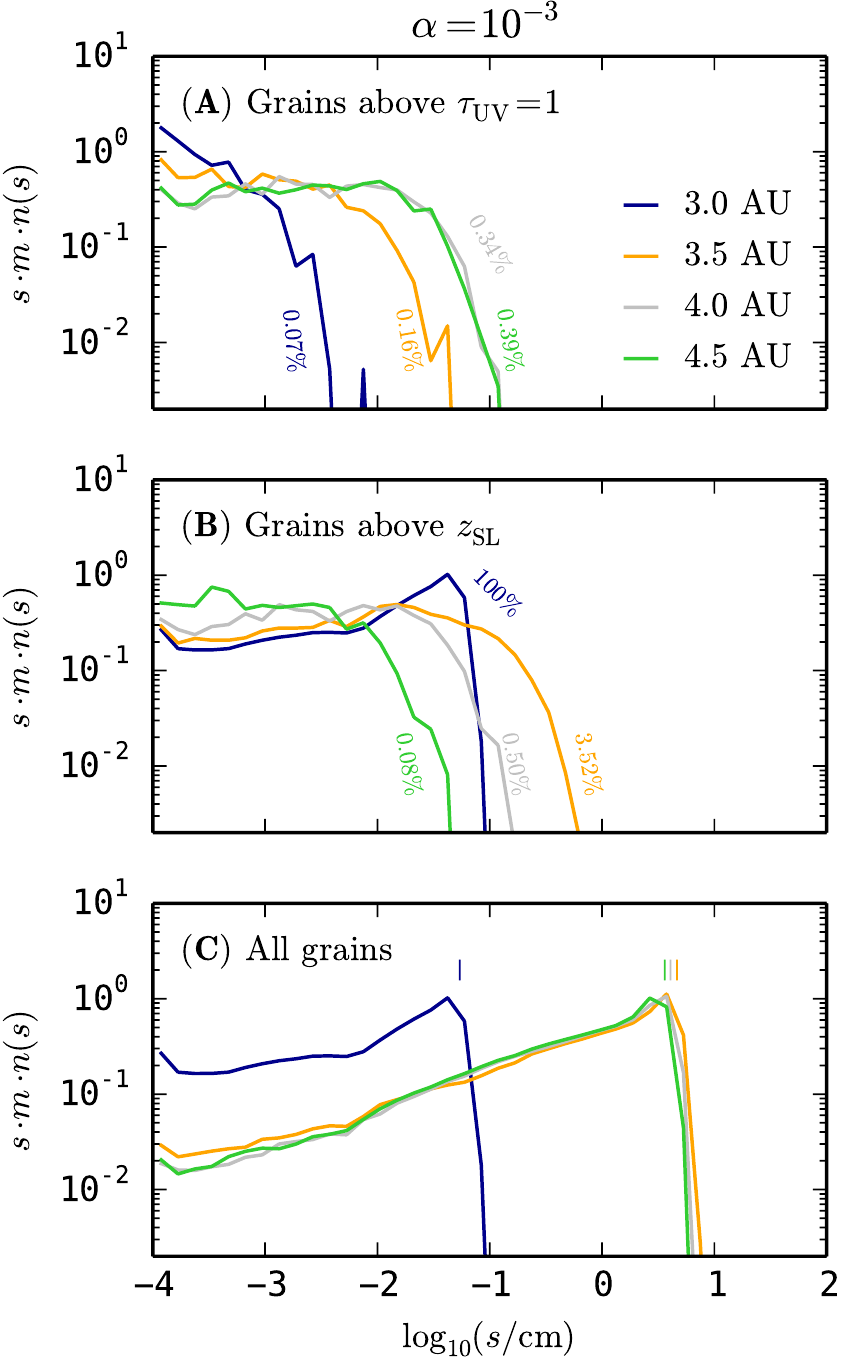}
\caption{Normalized dust size distributions in different vertical regions. (A): Only grains above the $\tau_\mathrm{UV}=1$ surface. The percentages show the total (i.e., size-integrated) dust mass that is present in this region relative to the total dust mass in the column. (B): Only grains above the vertical snowline $z_\mathrm{SL}$. (C): All grains. The vertical markers indicate $s_\mathrm{frag}^\mathrm{(Ep)}$ (Eq. \ref{eq:s_frag}) using $v_\mathrm{frag}=1\mathrm{~m/s}$ (for the $3\mathrm{~AU}$ case) and $v_\mathrm{frag}=10\mathrm{~m/s}$ (the other cases).}
\label{fig:namas}
\end{figure}

\subsection{Dust distributions}\label{sec:nama}
Figure \ref{fig:namas} shows the mass-weighted dust size distributions in different vertical regions of the column: the region above $z_{\tau=1}$, the region above $z_\mathrm{SL}$, and the entire column. The area under the curves is normalized to 1, and by plotting the quantity $s \cdot m \cdot n(s)$, the peak of the distribution shows where most of the solid mass is located. As in Sect. \ref{sec:IC}, we neglect the contribution of the ice mantles on particle mass and size\footnote{This results in an error in the particle size of $s^*/s = (1+f_\ice \rho_\bullet / \rho_\ice)^{1/3} $, where $s^*$ is the corrected particle size that takes into account the presence of an ice mantle with density $\rho_\mathrm{ice}=1\mathrm{~g/cm^3}$. For $f_\mathrm{ice} \sim 1$ (see Sect. \ref{sec:icerock}), this results in $s^*/s \sim 1.5$, which is small compared to the size range of the dust population.}. The vertically integrated distributions (bottom panel) closely resemble the coagulation/fragmentation steady-state distributions one would expect \citep[e.g.,][]{birnstiel2011}, with most mass concentrated in the peak close to the maximum size, which, for turbulence-induced fragmentation in the Epstein regime equals \citep{birnstiel2012}
\begin{equation}\label{eq:s_frag}
s_\mathrm{frag}^\mathrm{(Ep)} \approx \frac{2}{3\pi} \frac{\Sigma_\mathrm{g}}{\alpha \rho_\bullet} \left( \frac{v_\mathrm{frag}}{c_s} \right)^2.
\end{equation}
In Fig. \ref{fig:namas}C, there is a sharp transition between $3$ and $3.5\mathrm{~AU}$, as we go from ice-free grains inside the snowline to sticky, ice-rich aggregates (at least in the midplane) outside of ${\approx}3.2\mathrm{~AU}$. The difference in the maximum size is two orders of magnitude, as expected from Eqs. \ref{eq:v_frag} and \ref{eq:s_frag}. This sharp change in the vertically integrated size distribution around the radial snowline is expected to result in an observable jump in the spectral index, detectable with ALMA \citep{banzatti2015,cieza2016}. 

Since we have spatial information on the dust particles, we can also study the variation of the size distribution with height. Figure \ref{fig:namas}B shows the particle distribution for the grains that are located above the vertical snowline $z_\mathrm{SL}$. To highlight their shape, the curves have again been normalized, but the percentages indicate the total mass (relative to the total amount of dust in the column) in these populations. As we move from $r=3.5$ to $4.5\mathrm{~AU}$, the snowline moves up (e.g., Table \ref{tab:results}), resulting in a decrease in the mass fraction of dust above this surface (from 3.5\% to 0.08\%) and a decrease in the size of the largest particles that make it up there. Similarly, Fig. \ref{fig:namas}(A) shows the distribution above $z_{\tau_\uv=1}$. This would be the distribution that would be seen in short-wavelength observations of a nearly face-on disk. The $\tau_\uv=1$ surface lies relatively high for the $r=3\mathrm{~AU}$ case because the efficient fragmentation of ice-free grains results in a high abundance of small grains with a high surface/mass ratio (Eq. \ref{eq:tau_i}). As a result, the grains that are visible at higher $z$ are the smaller and less abundant then for the simulations at larger radii. In summary, depending on which vertical region one is interested in, the shape of the dust distribution can deviate significantly from the one in the midplane (essentially the distribution of Fig. \ref{fig:namas}C).

\subsection{Ice/rock ratios of solids}\label{sec:icerock}
\begin{figure}[t]
\centering
\includegraphics[clip=,width=.95\linewidth]{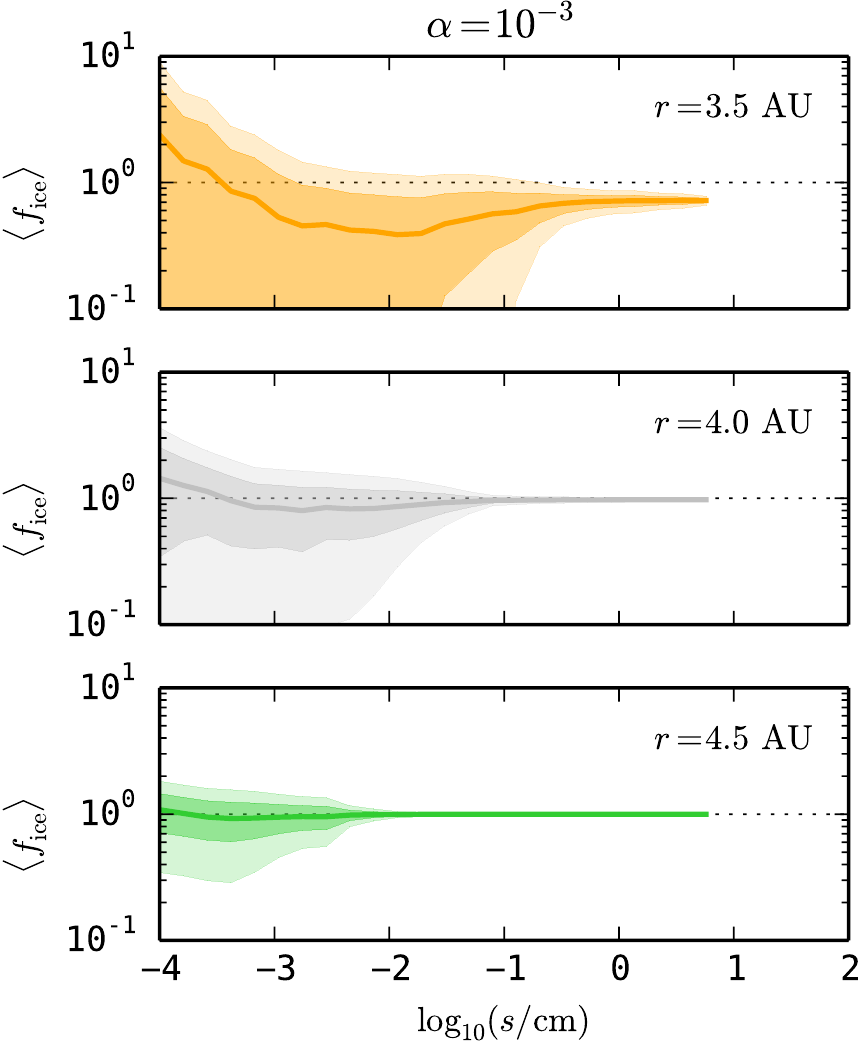}
\caption{Average ice-to-rock ratio as a function of dust particle size at the end of the simulations using $\alpha=10^{-3}$. The shaded areas indicate deviations from the average by $1$ and $2\sigma$, respectively, and the dotted horizontal line corresponds to $f_\ice = \Sigma_{\water} / \Sigma_\mathrm{d} = 1$ (see text).}
\label{fig:ice}
\end{figure}

For the cases outside of $r=3\mathrm{~AU}$, virtually all grains in the midplane are covered in water ice. Figures \ref{fig:35AU_NT} and \ref{fig:4AU_NT} show that the highest ice-rock mass ratios are found in the smallest grains, with sizes between $1-10\mathrm{~\mu m}$. This is to be expected, because the timescale for doubling ones mass purely by capturing water molecules scales as $s^3 / (\derd n_\ice / \derd t) \propto s^{-1}$ (see Eq. \ref{eq:dndt}) and is shorter for smaller particles. Figure \ref{fig:ice} shows the mean value and standard deviation of the ice/rock ratio as a function of particle size at different locations in the disk. In all cases, $\langle f_\ice \rangle \sim 1$ for the largest grains in the midplane, reflecting the fact that there is about as much water as dust in our column, i.e., $\Sigma_{\water} = \Sigma_\mathrm{d}$. The spread in the ice/rock ratio of the largest bodies is also small, which we attribute to the many collisions these particles see, which act to average out ice/rock ratios if particles stay in the same region. As we focus on smaller particles, the spread in $f_\ice$ becomes larger as these grains can cross the vertical snowline more readily.

\subsection{Vapor depletion in disk atmosphere}\label{sec:depletion}
Over the course of the simulations shown in Figs. \ref{fig:35AU_NT} and \ref{fig:4AU_NT}, the water abundance in the upper layers of the disk drops by a factor of ${\sim}2.5$ and $11$ (i.e., $1/\Delta_\mathrm{atm}$ in Table \ref{tab:results}), respectively, reaching a steady state value at the end of the 10 mixing times that were simulated. At $4.5\mathrm{~AU}$ the depletion is as high as a factor ${\sim}30$. The mechanism behind this depletion can be understood in terms of the `vertical cold finger' effect \citep{meijerink2009}, and is related to the transport and dynamics of water vapor and ice. Because of condensation there is always less water vapor (relative to $\hy$) in the midplane, resulting in a net flux of water vapor from the atmosphere down to the midplane. At the same time, there is turbulent mixing of ice-covered grains from the midplane back to the atmosphere, where these grains release their volatiles and replenish the water vapor. When all the grains are small, these two fluxes (vapor diffusing down \& ice being mixing back up) are balanced, and the water abundance above $z_\mathrm{SL}$ does not change\footnote{When we run our simulations without growth, we see no variation of the vapor abundance in time, and the snowline stays at $z_\mathrm{SL}^0$.}. However, when grains start to grow to sizes where settling becomes important, the flux of water ice being mixed from the midplane to $z>z_\mathrm{SL}$ diminishes. Then, the diffusion of water vapor wins, and drains the atmosphere of its water vapor on a timescale comparable to the mixing timescale $t_D$. The cold finger effect is most efficient when there are not a lot of small grains around, in which case it is easier for water molecules to travel close to the midplane before freezing out (see \citealt{monga2015} and Sect. \ref{sec:diffcon}). However, even if water molecules preferentially freeze out onto the smallest dust grains, water ice can still be transported to the largest, settled grains through sticking collisions. While the water vapor abundance can drop more than an order of magnitude locally above the snowline, the change in the vertically integrated water vapor to gas ratio (i.e., $2 \int_0^{\infty} \rho_\water(z) \intd z / \Sigma_\mathrm{g}$) is less dramatic because a significant portion of the water mass is in the region below the snowline where $\rho_\water=\rho^K_\mathrm{sat}$ at all times. Table \ref{tab:results} shows that the changes between the initial and final integrated water-to-gas ratio $\Sigma_\water / \Sigma_\mathrm{g}$ are of the order of $10-30\%$.
 
%: $2 \int_0^{\infty} \rho_\water(z) \intd z / \Sigma_\mathrm{g}$

\begin{deluxetable*}{c c  c  c | c c c c c c c }
\tabletypesize{\scriptsize}
%\rotate  
\tablecaption{Summary of model results.}
%\tablewidth{0pt}

\tablehead{
\colhead{$r$} & \colhead{$h_\mathrm{g}$} & \colhead{$\alpha$}  & \colhead{$t_D$} & \colhead{$s_\mathrm{max}$} & \colhead{$(\rho_\mathrm{d}/\rho_\mathrm{g})_\mathrm{mid}$}  & \colhead{$(n_\water/n_\hy)_\mathrm{atm}^\mathrm{(a)}$} & \colhead{$\Delta_\mathrm{atm}^\mathrm{(b)}$} &\colhead{$\Sigma_\water/\Sigma_\mathrm{g}^\mathrm{(c)}$}  &\colhead{$z_\mathrm{SL}$}  &  \colhead{$z_{\tau_\uv=\mu_0}$}
\\
\colhead{$\mathrm{[AU]}$} & \colhead{$\mathrm{[AU]}$} &  & \colhead{$\mathrm{[kyr]}$} & \colhead{$\mathrm{[cm]}$} & & &  &   &\colhead{$\mathrm{[AU]}$}  &  \colhead{$\mathrm{[AU]}$}
}
%\footnote{}

\startdata

\multicolumn{4}{c|}{\emph{Model parameters}} & \multicolumn{7}{c}{\emph{Initial conditions$^\mathrm{(d)}$}} \\
$3.0$ & $0.13$ & $-$ & $-$ & $10^{-4}$ & $5.0\times10^{-3}$ & $5.6\times10^{-4}$ & $-$ & $5.0\times10^{-3}$ & $0$ & ${>}0.53$ \\
$3.5$ & $0.16$ & $-$ & $-$ & $10^{-4}$ & $5.0\times10^{-3}$ & $5.6\times10^{-4}$ & $-$ & $1.8\times10^{-3}$ & $0.29$ & ${>}0.63$ \\
$4.0$ & $0.19$ & $-$ & $-$ & $10^{-4}$ & $5.0\times10^{-3}$ & $5.6\times10^{-4}$ & $-$ & $2.5\times10^{-4}$ & $0.53$ & ${>}0.75$ \\
$4.5$ & $0.22$ & $-$ & $-$ & $10^{-4}$ & $5.0\times10^{-3}$ & $5.6\times10^{-4}$ & $-$ & $3.3\times10^{-5}$ & $0.77$ & ${>}0.87$ \\

\multicolumn{4}{c|}{} & \multicolumn{7}{c}{\emph{After 10 mixing times}} \\
$3.0$ & $0.13$ & $10^{-3}$ & $0.8$ & $0.09$ & $5.6\times10^{-3}$ & $5.6\times10^{-4}$ & $1$ &$5.0\times10^{-3}$ & $0$ & $0.46$ \\
$3.5$ & $0.16$ & $10^{-3}$ & $1.0$ & $6.95$ & $2.1\times10^{-2}$ & $2.1\times10^{-4}$ & $0.38$ &$1.4\times10^{-3}$ & $0.17$ & $0.50$ \\
$4.0$ & $0.19$ & $10^{-3}$ & $1.3$ & $6.20$ & $2.2\times10^{-2}$ & $5.2\times10^{-5}$ & $0.09$ &$1.8\times10^{-4}$ & $0.31$ & $0.53$ \\
$4.5$ & $0.22$ & $10^{-3}$ & $1.5$ & $5.79$ & $2.3\times10^{-2}$ & $1.7\times10^{-5}$ & $0.03$ &$2.4\times10^{-5}$ & $0.46$ & $0.55$ \\ 

\multicolumn{4}{c|}{} & \multicolumn{7}{c}{\emph{After 10 mixing times}} \\

$3.0$ & $0.13$ & $5\times10^{-4}$ & $1.6$ & $0.17$ & $7.0\times10^{-3}$ & $5.6\times10^{-4}$ & $1$ &$5.0\times10^{-3}$ & $0$ & $0.43$ \\
$3.5$ & $0.16$ & $5\times10^{-4}$ & $2.1$ & $13.3$ & $3.8\times10^{-2}$ & $1.9\times10^{-4}$ & $0.33$ &$1.3\times10^{-3}$ & $0.16$ & $0.44$ \\
$4.0$ & $0.19$ & $5\times10^{-4}$ & $2.5$ & $12.3$ & $4.0\times10^{-2}$ & $4.2\times10^{-5}$ & $0.08$ &$1.7\times10^{-4}$ & $0.29$ & $0.48$ \\
$4.5$ & $0.22$ & $5\times10^{-4}$ & $3.0$ & $11.2$ & $4.1\times10^{-2}$ & $1.4\times10^{-5}$ & $0.02$ &$2.3\times10^{-5}$ & $0.45$ & $0.55$ 
\enddata
\tablecomments{The uncertainties, estimated by looking at the variation over the last $2t_D$ of the simulations, are ${\sim}3\%$ for the maximum grain size, ${\sim}1\%$ for the dust-to-gas ratio and atmospheric and integrated water abundance, and ${\sim}0.01\mathrm{~AU}$ and ${\sim}0.03\mathrm{~AU}$ for the snowline and $\tau_\uv=\mu_0$ surface locations, respectively.}
\tablenotetext{(a)}{The water vapor abundance in the well-mixed region above the snowline.}
\tablenotetext{(b)}{The atmospheric depletion factor $\Delta_\mathrm{atm}$ is defined as the ratio of the final to initial water vapor abundance in the disk atmosphere.}
\tablenotetext{(c)}{The water vapor surface density relative to $\Sigma_\mathrm{g}$ (see text). The initial value is calculated by first assuming that all the water is vapor and well-mixed with the gas, and then allowing it too freeze out beneath the vertical snowline (see Section \ref{sec:IC}).}
\tablenotetext{(d)}{The calculated initial snowline and $\tau_\uv=\mu_0$ surface locations are consistent with Eqs. \ref{eq:ZSL_0} and \ref{eq:ZTAU_0}.}

\label{tab:results}    
\end{deluxetable*}

\subsection{Location of the vertical snowline}\label{sec:location}
\begin{figure}[t]
\centering
\includegraphics[clip=,width=1.\linewidth]{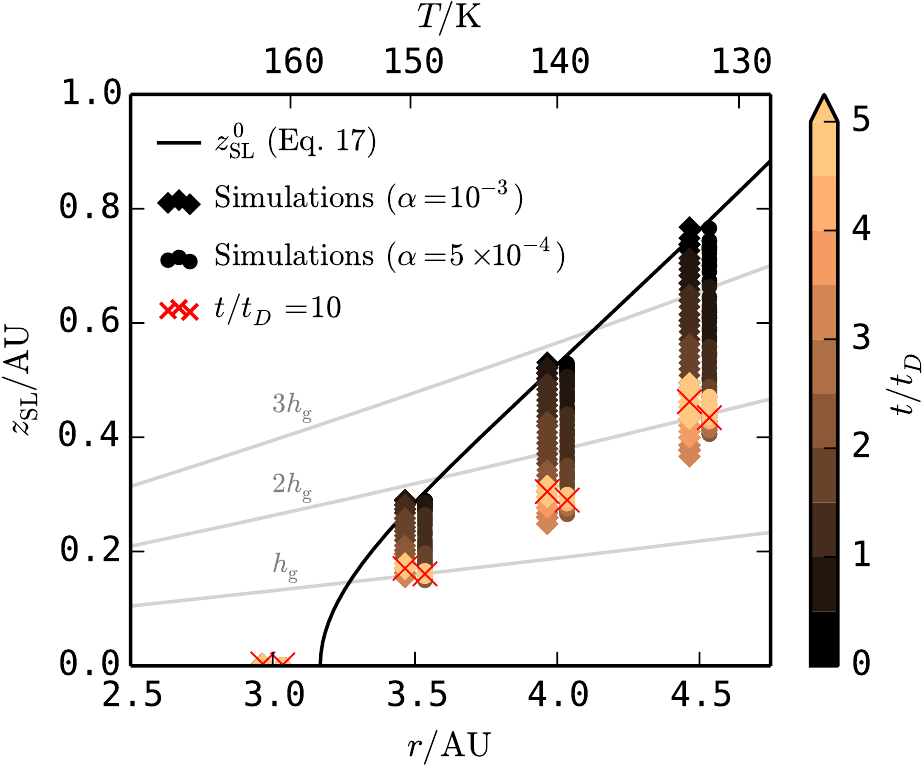}
\caption{Temporal evolution of the location of the vertical snowline at different radii and temperatures in the disk for two different values of $\alpha$. Red crosses indicate steady-state locations after 10 mixing times.}
\label{fig:zsl}
\end{figure}

The removal of water vapor from the disk atmosphere discussed in Sect. \ref{sec:depletion} has consequences for the location of the vertical snowline. Basically, the snowline is located at that height where the partial pressure of water equals the saturated pressure, $P_\water = P^K_\mathrm{sat}$, where $P^K_\mathrm{sat}$ is constant in the isothermal column. When the water mixing ratio drops, the snowline will shift to a region where the total gas pressure is higher, i.e., closer to the midplane. In Figs. \ref{fig:35AU_NT} and \ref{fig:4AU_NT}, the snowline moves down by $0.12$ and $0.22\mathrm{~AU}$, respectively (Table \ref{tab:results}), corresponding to about 1 scale-height at these locations. Fig. \ref{fig:zsl} shows the time evolution of the location of the vertical snowline between $r=3{-}4.5\mathrm{~AU}$. Initially, all simulations start with the snowline at $z_\mathrm{SL}^0$ (Eq. \ref{eq:ZSL_0}), but after the growth and settling of solids, a few mixing times are enough to lower $z_\mathrm{SL}$ by one, and in some cases almost two, scale-heights.

It is important to note that the vapor distribution above the snowline reaches a finite steady-state value, rather than continued depletion to $n_\water=0$. This can again be understood by thinking about the transport in terms of vapor diffusing down and ice-rich solids mixing up (specifically, through the location of $z_\mathrm{SL}$). First, as more water freezes out onto grains in the isolated column, the ice-rock ratio of the solids increases. Second, as the vertical snowline moves down, it reaches regions of increasing dust-to-gas ratio. These two effects result in the ice-flux (essentially the product of the flux of dust particles times their ice-rock ratios) \emph{increasing} as $z_\mathrm{SL}$ approaches the midplane, resulting in the behavior observed in Fig. \ref{fig:zsl}.

% We discuss the effect of changing the turbulence strength $\alpha$ on the steady-state value in Sect. \ref{sec:alpha}.

Lastly, Fig. \ref{fig:zsl} reveals that the snowline location reaches a minimum after a couple of mixing times, before increasing slightly to reach the steady-state value. This minimum corresponds to the period in the dust evolution just before small grains start to be replenished by collisions at $v_\mathrm{rel}>v_\mathrm{frag}$, when the abundance of small grains -- the most efficient transporters of icy material to the atmosphere -- is at an all-time low (this phase is visible in the 4th panel of Fig. \ref{fig:4AU_NT}).

%Typically, the degree of settling for a grain is governed by the ratio $\mathrm{St}/\alpha$, with a larger ratio indicating more settling. Thus, reducing $\alpha$ makes settling more efficient. 

\subsection{Effect of varying $\alpha$}\label{sec:alpha}
\begin{figure*}[t]
\centering
\includegraphics[clip=,width=.95\linewidth]{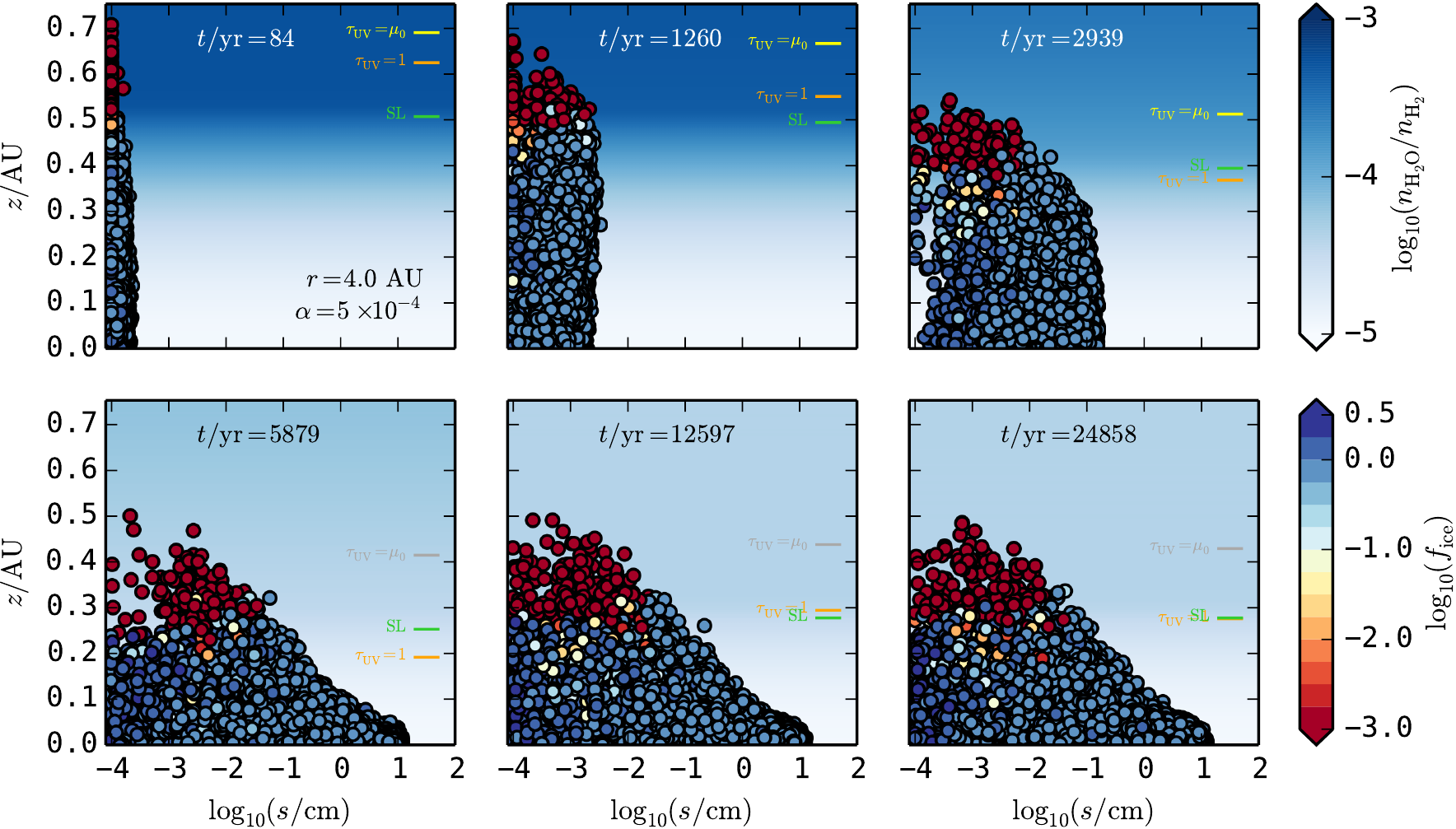}
\caption{Similar to Fig. \ref{fig:35AU_NT} but for a reduced turbulence $\alpha=5\times10^{-4}$. The weaker turbulence allows grains to grow to larger sizes while the slower vertical mixing results in grain settling becoming more important.}
\label{fig:4AU_WT}
\end{figure*}

Here we briefly discuss how the effects described in the previous sections change when the turbulence strength is reduced to $\alpha=5\times 10^{-4}$. Figure \ref{fig:4AU_WT} shows a simulation with (nearly) identical initial conditions to Fig. \ref{fig:4AU_NT}, but assuming a weaker turbulence. Several things stand out: First, the timescales are longer (the mixing time scales as $t_D \propto 1/\alpha$). Second, settling is more effective at keeping grains confined to the midplane. Finally, dust particles grow to larger sizes when turbulence is weaker because the relative collision velocity is decreased. Comparing the maximum sizes that are listed in Table \ref{tab:results}, we see that the lower turbulence results in maximum grain sizes that are a factor or ${\sim}2$ larger. This is in agreement with Eq. \ref{eq:s_frag}, which indicates $s_\mathrm{frag}\propto \alpha^{-1}$.

Generally, the fact that particles can grow larger and settle more readily results in the vapor being depleted more effectively in the $\alpha=5\times10^{-4}$ case. Comparing the resulting water vapor abundances in Table \ref{tab:results}, we see indeed that the water abundance in the atmosphere is between $10{-}50\%$ lower at the end of the simulations with a weaker turbulence. As a result, the snowline also moves further down, as illustrated in Fig. \ref{fig:zsl} where $z_\mathrm{SL}(t)$ is plotted for both values of $\alpha$. However, while the steady-state atmospheric vapor abundance and snowline location are lower in the $\alpha=5\times10^{-4}$ simulations, the time it takes the column to reach that steady-state is longer because of the increased mixing timescale (listed in Table \ref{tab:results}) as well as the fact that coagulation is slower for reduced relative collision velocities.

Simulating columns with a much weaker turbulence is challenging because of the very small relative scale-height of the largest grains, which results in a very small time-step in Eq. \ref{eq:dt_global}. In addition, a weaker turbulence results in a broader size distribution and therefore needs more representative particles to accurately resolve it. Several elegant methods have been proposed to alleviate the computational costs in such cases \citep[e.g.,][]{charnoz2012, drazkowska2013}, though it may not be straightforward to use them in the context of simultaneous dust and vapor evolution.

\subsection{Impact on gas-phase C/O ratio}
The gas-phase $\cor$ ratio heavily influences the chemistry taking place inside protoplanetary disks and the effect of the radial snowlines of $\water$, $\co$, and $\co_2$ on the $\cor$ ratio in the gas and solids in the midplane has been investigated by \citet{oberg2011} and recently, in the context of radial drift, by \citet{piso2015}. The vertically resolved models introduced in this work allow for the study of the vertical and temporal variations of the gas-phase $\cor$ ratio across the vertical snowline. Generally, the removal of oxygen-bearing molecules (water in this case) from the disk atmosphere will increase the $\cor$ ratio in these regions as most carbon is in forms that are more volatile than water. By making some basic assumptions about the carbon abundances in the gas and refractory solids, we can estimate the magnitude of this effect.

With our models being located inside the radial snowline of $\co$, we assume the gas-phase carbon is mainly dominated by carbon-monoxide, and assume a vertically constant abundance of $n_\co/n_\hy = 4\times10^{-4}$ \citep[e.g.,][]{oberg2011}. For the initial conditions in our simulations, we can calculate the initial C/O ratio above the snowline to be $(\cor)_\mathrm{atm} = 0.4$. Grain growth and settling alone result in $(\cor)_\mathrm{atm} = 0.9$ at the end of the simulation of Fig. \ref{fig:4AU_NT}, and for the largest depletions found in Table \ref{tab:results} ($\alpha=5\times10^{-4}$ at $4.5\mathrm{~AU}$) we find $(\cor)_\mathrm{atm} \approx 1$. If a considerable fraction of the available gas-phase carbon is present in the form of oxygen-poor molecules (e.g., $\mathrm{CH_4}$, atomic carbon), $\cor>1$ could well be reached above the vertical snowline. For the solids in the midplane we expect the opposite effect, as the extra water that is added to the dust grains in the midplane will \emph{decrease} the C/O ratio of these particles. As the ice-to-rock ratio varies with height, but also with particle size (e.g., Fig. \ref{fig:ice}), we expect the $\cor$ in the solids vary significantly even at a single disk location.

In conclusion, our vertically resolved models indicate that both the gas-phase and solid-phase $\cor$ ratio display significant variations at a single disk radius. At larger disk radii, outside the radial snowlines of important carbon carriers like $\co$, this picture becomes more complex. Observational studies are starting to pin down volatile abundances in these parts \citep{du2015, kama2016, bergin2016}, finding volatiles to be depleted from the gas and possibly pointing toward a mechanism similar to the one described in Sect. \ref{sec:depletion}. To study these regions through numerical simulations however, the radial drift of solids and photodesorption (see Sect. \ref{sec:PD}) will have to be included \citep{piso2015,cleeves2016}.

\section{Discussion}\label{sec:disc}
The goal of this work has been to show that, even for the relatively simple case of an isolated, isothermal column and standard assumptions for the dust evolution, dynamical effects play an important role in setting the vertical abundances of dust, ice, and water vapor. In that light, we have neglected several complex and more subtle processes that can influence the observed behavior. Here, we briefly discuss some of those effects.

\subsection{Dust particle model}\label{sec:dpmodel}
Throughout this work we have assumed the dust grains are well-described as compact and (roughly) spherical aggregates. In reality, coagulation is expected to form aggregates with a complex internal structure, whose (average) porosity is the result of their growth history and potentially non-collisional compaction mechanisms \citep{ormel2007,okuzumi2012,kataoka2013c,krijt2015}. A high porosity is expected to influence collisional outcomes \citep{dominiktielens1997,wada2011}, typically increasing the fragmentation velocity as porous grains are better at dissipating collisional energy. Porous aggregates are often described as fractal-like structures, with a mass-radius relation $m\propto s^x$ with $2<x<3$ \citep{ormel2007,suyama2012}. Thus, the surface-to-mass ratio does not decrease as rapidly with increasing mass, changing the collisional cross sections and aerodynamic properties of growing aggregates. Specifically, the physical particle sizes at which settling and fragmentation start to dominate will increase, while the maximum Stokes number that can be reached in fragmentation-limited growth is independent of the particle's internal density \citep{birnstiel2012}. 

Another assumption of our model is that the refractory core of the aggregate stays intact when the ice mantle is lost. This is not necessarily the case, in particular when the dust particles are better described as porous aggregates of individual ice-covered monomers, in which case crossing a snowline can result in the aggregate disintegrating \citep{saito2011}. Such non-collisional disruption could lead to an increase in the abundance of small refractory fragments just above the snowline, possibly changing the optical depth locally.

%This picture is complicated by a process known as sintering. Driven by variations in the local radius of curvature, sintering refers to water molecules moving from regions with a positive radius of curvature to a negative radius of curvature. Initially, this results in the strengthening of monomer-monomer bonds, which makes the aggregate more brittle and less efficient at dissipating collisional energy , especially in regions close to the snowlines of important volatile species \citep{sirono2011,okuzumi2016}. In more extreme cases, sintering can result in the spontaneous break-up of aggregates, increasing the prevalence of small fragments \citep{sirono2011b}. 

In deriving Eq. \ref{eq:dndt} we have assumed that the condensation and sublimation of water molecules are processes that do not depend on the size or composition of the grain. In reality, the desorption energies for water molecules also depend on the molecular composition of the grain surface. Compared to a water ice surface, the desorption energy is lower for a carbon surface and higher for a silicate surface \citep{papoular2005,cuppen2007,goumans2009}. Thus, depending on the composition of the refractory cores in our model, the formation of the first monolayer of water-ice can then be somewhat faster (for silicate grains) or slower (for carbonaceous grains) compared to what we have assumed here. With most grains in our simulations being covered in many monolayers, the impact of this effect on our simulations is small.

Because we picture the aggregates as collections of microscopic monomers (Fig. \ref{fig:schematic_1}), we assume there are no differences in surface curvature between aggregates of different sizes. While $K\sim1/s_\bullet$ is a good approximation for the area-averaged curvature, the small regions where monomers are in contact with each other will have a negative radius of curvature. Vapor will preferentially freeze-out around these contacts, causing the inter-monomer bonds to harden and the aggregate to become more brittle. This process, known as sintering, can influence the aggregate's collisional behavior \citep{sirono1999}, in particular in regions where the sintering timescale is short compared to the collision timescale \citep{sirono2011,okuzumi2016}.

%\todo{In addition ... (curvature affected by water freezing out?)}

Developing a more physical model for the evolution of the internal structure (i.e., porosity) of the aggregates in the presence of grain-grain collisions, sintering, and condensation/sublimation will be the subject of future work. While including these effects are expected to impact the shape of the dust and ice distributions, the loss of water vapor from the disk atmosphere as described in Sect. \ref{sec:depletion} is inevitable after grain growth takes place and aggregates that experience some degree of settling, and are capable of accumulating an ice mantle, are being formed. Furthermore, the formation of larger solids that are able to overcome the meter-size barrier via, e.g., the streaming instability or dust trapping, will trap ices. Depending on the size of these bodies they might trap ices even when environmental temperatures are above the sublimation temperature of water ice. Thus, volatile depletion is an inevitable consequence of grain growth/planetesimal formation.

% This picture is somewhat idealized because we know these processes can depend on the (local) radius of curvature \citep{sirono2011b,kuroiwa2011} and the composition of the grain surface \citep[e.g.,][]{cuppen2007}.

%The equilibrium vapor pressure on a curved surface can be approximated as $P^*_\mathrm{sat} = P_\mathrm{sat} (1+ K \gamma v / k_\mathrm{B} T)$, with $\gamma=69\mathrm{~erg/cm^2}$ the surface energy of water ice, $v=3.3\times10^{-23}\mathrm{~cm^3}$ the volume of a water molecule, and $K\sim 1/s$ the radius of curvature \citep{sirono2011}. At $T=140\mathrm{~K}$, this results in a fractional change in the equilibrium vapor pressure of $0.1\%$ for spherical micron-size grains, and $10^{-5}\%$ for cm-size grains, indicating a minor impact on $\dot{n}_\ice$ for the cases considered here.

%\subsection{Porosity and sintering}

%

\subsection{Photo-desorption}\label{sec:PD}
In our calculations, we have ignored the process of photo-desorption, in which incident UV photons remove water molecules from grain surfaces. The resulting loss rate of water molecules (in $\mathrm{g~cm^{-2}~s^{-1}}$) can be written as $F_\mathrm{PD} =  m_\water F_0 Y e^{-\tau_\uv / \mu_0} $ \citep[e.g.,][]{dalessio1998,ciesla2014}, with $Y=10^{-3}$ the typical yield per UV photon \citep{oberg2009} and $F_0$ the stellar UV flux at this specific radial location. Since the snowline is usually well shielded from UV photons in our simulations (i.e., $z_\mathrm{SL} < z_{\tau_\uv=\mu_0}$) it is safe to neglect photo-desorption. For example, near the snowline location at the end of the simulation shown in Fig. \ref{fig:4AU_NT}, and assuming an $F_0$ of $10^4$ times the interstellar radiation field of $G_0=10^8\mathrm{~cm^{-2}~s^{-1}}$ \citep{habing1968}, we obtain $F_\mathrm{sub} \approx F_\mathrm{con} \approx 10^{12} F_\mathrm{PD}$, justifying our choice of neglecting photo-desorption in Eq. \ref{eq:dndt}. At larger radii however ($r>20\mathrm{~AU}$), where gas densities and temperatures are lower, photo-desorption can play an important role \citep{vandishoeck2014}, and will have to be included. With the vertical profile of the optical depth readily available in our models, we will be able to include photo-desorption self-consistently in future calculations.

\subsection{Diffusion and condensation beneath the snowline}\label{sec:diffcon}
Our method assumes the gas-phase water is well-mixed within a volume element $\mathcal{L}^3$. If the condensation timescale for the water molecules that are moving down and crossing the snowline is very short however, this assumption can break down when these water molecules freeze out in a region just below $z_\mathrm{SL}$, as the thickness of this region is potentially ${\ll}\mathcal{L}$. In that case, our assumption would lead us to overestimate the efficiency of water vapor transport to regions deeper in the disk. Assuming that the (local) dust population can be characterized by a single grain size $s$, the timescale for a single water molecule to freeze out equals $t_\mathrm{con} \simeq s \rho_\bullet / v_\mathrm{th} \rho_\mathrm{d}$. This timescale can be used to calculate a lengthscale $l_\mathrm{con} \sim \sqrt{ 2 D_\mathrm{g} t_\mathrm{con}}$, assuming diffusion is responsible for the motion of the water vapor molecule. Plugging in $s=1\mathrm{~\mu m}$ and a local dust density of small grains that corresponds to 1\% of the total dust mass (see Fig. \ref{fig:namas}C), we obtain $ l_\mathrm{con} / h_\mathrm{g} \sim 10^{-2} - 10^{-1}$, varying somewhat with location (height above the midplane) and turbulence strength. Before the dust growth/fragmentation steady-state is reached, there is a period of time where the abundance of small grains is reduced, during which $l_\mathrm{con}$ is larger. Thus, in general $l_\mathrm{con} \gtrsim \mathcal{L}$ (see Sect. \ref{sec:resolution}) and our assumption that the water vapor is well-mixed inside a single volume element is expected to hold.

\subsection{Vertical temperature structure}\label{sec:T}
For simplicity, we have assumed a vertically isothermal temperature profile in our simulations. Generally however, both stellar irradiation and accretional heating will contribute to the thermal balance in the disk \citep[e.g.,][]{armitage2010} and the vertical temperature profile -- as well as the difference in temperature between the midplane and atmosphere -- is determined by the strength of the turbulence and the vertical distribution of the opacity \citep{chiang1997,dalessio1998,dalessio2001}. Typically, stellar irradiation increases the temperature in the region at higher $z$ where the optical depth is below $\tau_\uv \sim \mu_0$ \citep[e.g.,][Fig. 4]{dalessio1998}, while viscous accretion increases the temperature in the disk midplane, although it is highly uncertain where in the nebula the dissipation of viscous energy takes place \citep{turner2014}. In our simulations, the snowline is always below $z_{\tau_\uv=\mu_0}$ (Table \ref{tab:results}).

A variable temperature may result in a $z$-dependent diffusivity, which can readily be included in Eqs. \ref{eq:diffusion_eq} and \ref{eq:D_g} \citep[see][]{ciesla2010}. Furthermore, the equilibrium water vapor pressure $P_\mathrm{sat}$ will become a function of disk height. While including a variable temperature will complicate things numerically, the vapor depletion as described in Sect. \ref{sec:depletion} will still occur as long as the midplane is cold enough for water molecules to freeze out -- as would be the case outside the radial snowline of a disk whose heating is dominated by irradiation -- although the magnitude of the snowline migration (e.g., Fig. \ref{fig:zsl}) could well be smaller if the vertical variation of the equilibrium vapor pressure is dominated by a (steep) temperature gradient rather than a change in bulk gas density.

%Alternatively, for high values of the mass accretion rate and $\alpha$, a warm midplane layer may exist where water ice does not form \citep{min2011}. In that case, the atmospheric vapor loss will be less severe as ices cannot be sequestered in the midplane regions as efficiently. 

\section{Summary}
We have developed a model that simultaneously simulates dust growth/fragmentation, dust grain dynamics, gas-grain chemistry, and vapor diffusion in a vertical column of a protoplanetary disk (Sect. \ref{sec:methods}). With this model, it is possible to connect the size distribution and water-content of the solids growing in the midplane to the vapor abundance and optical depth evolution in the disk atmosphere (e.g., Fig. \ref{fig:4AU_NT}). We have applied this method to study the impact dust coagulation has on the vertical distribution of solids, water ice, and water vapor, in the region around the radial snowline (Sect. \ref{sec:results}), finding that:

\begin{enumerate}

\item{Dust growth followed by gravitational settling can, even in the absence of radial migration of solids or growth to planetesimal sizes, deplete volatiles from the disk atmosphere, as hypothesized by \citet{meijerink2009}. The driving force behind this depletion is a reduction in the ability of the dust population to mix ice-covered material back up to the disk atmosphere (Sect. \ref{sec:depletion}).}

\item{These effects cause the water vapor abundance in the region above the vertical snowline to decrease by as much as a factor of ${\sim}50$, depending on the strength of the turbulence and the location in the disk (Table \ref{tab:results}). The water abundance does not go to zero however, because a small population of upward diffusing and ice-rich grains is maintained.}

\item{The removal of water from the vapor phase causes the vertical snowline to move closer to the midplane (Fig. \ref{fig:zsl} and Sect. \ref{sec:location}). The change in location of the vertical snowline depends on the turbulence but is typically of the order of a gas scale-height.}

\item{One consequence of the water vapor depletion in the disk atmosphere is that the $\cor$ ratio in the gas phase increases above the vertical snowline. At the same time, the $\cor$ ratio of the solid grains in the midplane is decreased.}

\end{enumerate}
Our simulations show how important dynamical effects (settling, turbulent mixing) can be and illustrate the need for vertically resolved and time-dependent modeling efforts. Improvements to the physical model describing the dust aggregates, the addition of other volatile species, and the inclusion of radial drift and a variable vertical temperature profile are the focus of follow-up work.

\acknowledgments
The authors thank M. Kama for fruitful discussions. This material is based upon work supported by the National Aeronautics and Space Administration under Agreement No. NNX15AD94G for the program ``Earths in Other Solar Systems.'' The results reported herein benefitted from collaborations and/or information exchange within NASA's Nexus for Exoplanet System Science (NExSS) research coordination network sponsored by NASA's Science Mission Directorate. The authors acknowledge funding from NASA grants NNX12AD59G and NNX14AG97G. EAB also acknowledges support from the National Science Foundation grant AST-1514670 and AST- 1344133 (INSPIRE) along with NASA XRP grant NNX16AB48G.

%\appendix

\bibliographystyle{aa}
\bibliography{refs}

\end{document}